\documentclass[twocolumn,showpacs,amsfonts,amsmath,amssymb,pra]{revtex4}
\usepackage{epsfig}
\usepackage{subfigure}
\usepackage{amsmath}
\usepackage{amssymb}
\usepackage{array}
\usepackage{pifont}
\begin{document}
\title{Absorption in atomic wires}%
\author{Jose M. Cerver\'o}%
   \email[To whom correspondence should be addressed:\\]{cervero@usal.es}%
    \homepage[\\Visit: ]{http://www.usal.es/~fnl/}
\author{Alberto Rodr\'{\i}guez}%
    \affiliation{F\'{\i}sica Te\'orica. Facultad de %
Ciencias. Universidad de Salamanca. 37008 Salamanca. Spain}%
\begin{abstract}%
      The transfer matrix formalism is implemented in the form of
the multiple collision technique to account for dissipative transmission processes
by using complex potentials in several models of atomic chains. The
absorption term is  rigorously treated to recover unitarity for the
non-hermitian hamiltonians. In contrast to other models of
parametrized scatterers we assemble explicit potentials
profiles in the form of delta arrays, P\"oschl-Teller holes and complex
Scarf potentials. The techniques developed  provide analytical expressions for the
scattering and absorption probabilities of arbitrarily long wires. The
approach  presented is suitable for modelling molecular aggregate
potentials and also supports new models of continuous disordered
systems. The results  obtained also suggest the possibility of using these
complex potentials within disordered wires to study the loss of coherence
in the electronic localization regime due to phase-breaking inelastic processes.
\end{abstract}
\maketitle
%%%%%%%%%%%%%%%%%%%%%%%%%%%%%%%%%%%%%%%%%%%%%%%%%%%%%%%%%%%%%%%%%%%%%%%%%%%%%
\section{Introduction}

The inelastic scattering processes occurring in mesoscopic samples as a
consequence of a finite non-zero temperature can noticeably  change the
coherent transport fingerprints of these structures. The worsening of 
electronic transmission due to such effects is expected but in some
situations the competition between the phase-breaking mechanisms and the quantum coherent
interferences can improve conductance in certain energetic
regimes. This is the case, for example, of disordered structures. This fact
 has attracted much attention  in the study and modelling of dissipative
transport in  one-dimensional structures. Interest is also prompted by 
experiments currently being carried out on real atomic chains \cite{vieira}.

A model of parametrized scatterers coupled through additional side channels to electron
reservoirs incorporating inelastic events was initially proposed by
B\"uttiker \cite{butti}, and much work has been done along this line \cite{bundle}. 
On the other hand, inelastic processes can be modelled by small absorptions
which in turn can be described by extending the nature of the quantum potentials to
the complex domain. The main purpose of this work is to include absorptive
processes by performing these complex extensions on previous quantum
wire models developed by the authors \cite{aro} and also on other atomic potentials.

The use of complex site energies and frequencies has already been 
considered in the study of electronic conductivity through one-dimensional 
 chains \cite{kramer,pataki}, but non-hermitian hamiltonians
have also been used to account for a large variety of phenomena, ranging from 
wave transport in absorbing media \cite{waves}, violation of the single
parameter scaling in one-dimensional absorbing systems \cite{deych}, 
appearance of exceptional points in scattering theory \cite{epscatt} and
quantum chaology \cite{epcaos}, description of vortex delocalization in
superconductors with a transverse Meissner effect \cite{nelson} and more
phenomenologically with nuclear optical potentials. Special mention is
required for the framework of $\mathcal{PT}$-symmetry \cite{pts}, where it is
possible to consider periodic wires under complex potentials showing  real
band spectra \cite{ptcer,ptperio}. There is nothing wrong in principle with
the use of non-hermitian hamiltonians as long as their properties
are controlled by a sufficient knowledge of the full spectrum. Indeed, 
renormalization group calculations have been carried out giving rise to
imaginary couplings as a result of quantum dressing of the classical  real
potentials \cite{renorma}. 

An interesting modern review on absorption in
quantum mechanics has appeared recently \cite{report} and we address the
interested reader to this publication and references therein. 

The paper is organized as follows. In Section \ref{sec:mct} we briefly
review the multiple collision technique based on the transfer matrix
method, and in Section \ref{sec:schro} we show how unitarity can be easily
restored in the presence of absorption and how the general unitarity condition
can be generalized accordingly. We then turn our attention to arrays of
delta potentials and calculate and draw the scattering and absorption
probabilities in Section \ref{sec:deltas}. In Section \ref{sec:posch},  
the P\"oschl-Teller potential is used to build atomic chains, and its
complex extension the complex Scarf potential is fully developed in Section
\ref{sec:Scarf}. The analytical scattering probabilities are shown for a
variety of composite potential profiles and the effect of the imaginary
parts on the transmission is analyzed. The calculations concerning exact
wave functions and analytical conditions are offered in two 
appendices. The paper ends with several concluding remarks.

%%%%%%%%%%%%%%%%%%%%%%%%%%%%%%%%%%%%%%%%%%%%%%%%%%%%%%%%%%%%%%%%%%%%%%%%%%%%%
\section{The multiple collision technique}
    \label{sec:mct}
    The time-independent scattering process in one dimension can be described using the
well known continuous transfer matrix method \cite{harrison},
\begin{equation}
    \begin{pmatrix} A_R \\ B_R \end{pmatrix}=
    \begin{pmatrix} M_{11} & M_{12} \\ M_{21} & M_{22} \end{pmatrix}
    \begin{pmatrix} A_L \\ B_L \end{pmatrix}
\end{equation}
where $A_L, B_L$$(A_R, B_R)$ mean the amplitudes of the asymptotic
travelling plane waves $e^{ikx},e^{-ikx}$ at the left (right) side of the potential. 
Whatever the nature of the potential is, real or complex, the transmission matrix always
verifies $\det\mathbf{M}=1$ as a consequence of the constant Wronskian
of the solutions of the Schr\"odinger equation.
The transmission and reflection amplitudes then read,
\begin{equation}
    t=\frac{1}{M_{22}} ;\quad r^L=-\frac{M_{21}}{M_{22}} ;\quad
    r^R=\frac{M_{12}}{M_{22}}
\end{equation}
where the superscripts $L,R$ stand for left and right incidence.
The insensitivity of the transmission amplitude to the incidence direction 
is a universal property that holds for all kind of potentials. However, the
reflectivity, although symmetric for real potentials, changes with the
incidence side for a complex one unless it is symmetric \cite{zafarpra}. 
The effect of a composition of $n$ different potentials can then be considered
as the product of their transmission matrices,
\begin{equation}
    \mathbb{M}=\mathbf{M}_n\ldots\mathbf{M}_2\mathbf{M}_1.
\end{equation}
The transmission matrix formalism is an important tool for the numerical
treatment of different problems.
An intuitive and general interpretation of the composition procedure can be
given in the following form. Consider two potentials $V_1(x), V_2(x)$
characterized by the scattering amplitudes
$t_1,r^L_1,r^R_1,t_2,r^L_2,r^R_2$ and 
joined at a certain point. Then, the scattering amplitudes of the composite potential can be
obtained by considering the coherence sum of all the multiple reflection processes that might
occur at the connection region,
\begin{subequations}
\label{eq:comp}
\begin{align}
    t &\equiv t_1\left\{\sum_{n=0}^\infty (r^L_2 r^R_1)^n\right\} t_2 =
    \frac{t_1 t_2}{1-r^L_2 r^R_1} \\
    r^L &\equiv r^L_1 +t_1 r^L_2 \left\{\sum_{n=0}^\infty (r^L_2 r^R_1)^n\right\}
    t_1 = r^L_1 +\frac{r^L_2 t_1^2}{1-r^L_2 r^R_1} \\
    r^R &\equiv r^R_2 + t_2 r^R_1 \left\{\sum_{n=0}^\infty (r^L_2
    r^R_1)^n\right\} t_2 = r^R_2 +\frac{r^R_1 t_2^2}{1-r^L_2 r^R_1}.
\end{align}
\end{subequations}
Replacing the scattering amplitudes with the
elements of the corresponding transmission matrices $\mathbf{M}_1,\mathbf{M}_2$, one can trivially 
check that in fact these last formulae are the equations of the product
$\mathbf{M}_2\mathbf{M}_1$. Thus, the composition rules given by
\eqref{eq:comp} are not restricted to the convergence interval of the
series $\sum_{n=0}^\infty (r^L_2 r^R_1)^n$. They provide an explicit
relation of the global scattering amplitudes  in terms of
the individual former ones and can be easily used recurrently for numerical
purposes.
This composition technique was first derived for a potential barrier
 \cite{beam} and has been used for designing absorbing potentials \cite{palao}.
%%%%%%%%%%%%%%%%%%%%%%%%%%%%%%%%%%%%%%%%%%%%%%%%%%%%%%%%%%%%%%%%%%%%%%%%%%%%%%%%%%%%%%%%
\section{The Schr\"odinger equation for a complex potential}
    \label{sec:schro}
Let us consider a one-dimensional complex potential of finite support
$V(x)=V_r(x)+i V_i(x)$ ($V(\pm\infty)=0$). For the stationary scattering states,
 the density of the current flux is proportional to the
imaginary part of the potential
\begin{equation}
    \label{eq:cont}
     \frac{dJ}{dx}=\frac{2}{\hbar} V_i(x) |\Psi(x)|^2
\end{equation}
where $J(x)$ is defined as,
\begin{equation}
    J(x)=\frac{\hbar}{2mi}\left(\Psi^*(x)\frac{d\Psi(x)}{dx}-\Psi(x)\frac{d\Psi^*(x)}{dx}\right).
\end{equation}
Therefore, in the presence of a non-vanishing $V_i(x)$ the unitarity
relation regarding the transmission and reflection probabilities
$T(E)+R(E)=1$ is no longer valid. One can still recover a pseudounitarity
relation by defining a quantity that accounts for the loss of flux in the
scattering process. Dealing with the asymptotic state
$\Psi^L_k(-\infty)=e^{ikx}+r^L(k)e^{-ikx}$, $\Psi^L_k(+\infty)
=t(k)e^{ikx}$, one can write the asymptotic values of the flux as,
\begin{align}
    J_{-\infty} &= \frac{\hbar k}{m}\left(1-R^L(k)\right) \\
    J_\infty &= \frac{\hbar k}{m}T(k)
\end{align}
yielding the relation,
\begin{equation}
    T(k)+R^L(k)+\frac{m}{\hbar k} \left( J_{-\infty}-J_\infty\right) = 1.
\end{equation} 
This latter equation remains the same for the right incidence case (with $R^R(k)$)
when the asymptotic state takes the form $\Psi^R_k(-\infty)=t(k)e^{-ikx}$, $\Psi^R_k(+\infty) =e^{-ikx}+r^R(k)e^{ikx}$.

Using eq. \eqref{eq:cont} the flux term reads,
\begin{align}
    A^{L,R}(k) & \equiv -\frac{2m}{\hbar^2 k}\int_{-\infty}^{\infty} V_i(x)
    |\Psi^{L,R}_k(x)|^2 dx \notag \\ &= 1-R^{L,R}(k)-T(k),
\end{align}
and it is usually understood as the probability of absorption
 \cite{zafarpra}. But $A(k)$ must be a positive defined quantity in order to be
strictly considered as a probability and this is not ensured by the
definition (unless $V_i(x)<0 \;\forall\,x$). The sign of $A(k)$ depends on both the changes in sign of the imaginary
part of the potential and the spatial distribution of the state. 
Although a negative value for $A(k)$ could be viewed as emission (because it
means a gain in the flux current) it also leads the transmitivity and the
reflectivity to attain anomalous values $T(k)>1, R(k)>1$, which are
difficult to interpret. Let us also note that the integral representation of
the absorption term is useless for practical purposes because to build the
correct expression of the state $\Psi^{L,R}_k(x)$ one needs to impose the
given asymptotic forms to the general solution of the Schr\"odinger equation, 
therefore obtaining the scattering amplitudes, so one cannot calculate 
the absorption probability without knowing $R(k)$ and $T(k)$.
%%%%%%%%%%%%%%%%%%%%%%%%%%%%%%%%%%%%%%%%%%%%%%%%%%%%%%%%%%%%%%%%%%%%%%%
\section{Scattering of a chain of delta potentials}
    \label{sec:deltas}
    Let us consider a potential constituted by a finite array of Dirac delta
distributions, each one with its own coupling $\alpha_i$ and equally spaced at
a distance $a$. This is probably  the simplest one-dimensional model 
imaginable, but in spite of its apparently simplicity it supports an
unexpected physical richness. It has been successfully used to model band
structure in a periodic quantum wire \cite{aro} and has proved its usefulness when
considering uncorrelated and correlated disorder structures
 \cite{aro,adame}, showing
interesting effects such as the fractality of the density of states and the
different localization regimes for the electronic states.

The global potential will be characterized by the arranged sequence of the
parameters $(a/a_i)$, where $a_i=\frac{\hbar^2}{m\alpha_i}$ means the
``effective range'' of the $i$th delta,  in the order they appear from
left to right. The transmission matrix for a delta potential preceded by a
zero potential zone of length $a$ reads,
\begin{equation}
    \label{eq:tmdelta}
    \mathbf{M}_j(k)=\begin{pmatrix} \left(1-\frac{i}{ka_j}\right)e^{ika} & 
    -\frac{i}{ka_j} e^{-ika} \\ \frac{i}{ka_j} e^{ika} & 
    \left(1+\frac{i}{ka_j}\right)e^{-ika} \end{pmatrix}.
\end{equation}
Considering a chain of $N$ different deltas and applying the composition rules to this type of matrices one finds that it is
possible to write a closed expression for the scattering amplitudes.
 They are given by,
\begin{subequations}
\label{eq:tyr}
\begin{align}
    t(k;a_1,\ldots,a_N) &= \frac{e^{iNka}}{f(k;a_1,\ldots,a_N)} \\
    r^L(k;a_1,\ldots,a_N) &= -\frac{g(k;a_1,\ldots,a_N)}{f(k;a_1,\ldots,a_N)}
\end{align}
\end{subequations}
with the definitions
\begin{widetext}
\begin{subequations}
\label{eq:fool}
\begin{align}
    f(k;a_1,\ldots,a_N) &= 1+\frac{i}{ka}\sum_{j=1}^N
    \left(\frac{a}{a_j}\right)+ \sum_{j=2}^N \left(\frac{i}{ka}\right)^j
    \left\{ \sum_\sigma \left[\left(\frac{a}{a_{\sigma_1}}\right) 
    \left(\frac{a}{a_{\sigma_2}}\right)\ldots
    \left(\frac{a}{a_{\sigma_j}}\right) \prod_{r=1}^{j-1}
    \left(1-e^{2ika(\sigma_{r+1}-\sigma_r)}\right)\right]\right\} \\
    g(k;a_1,\ldots,a_N) &= \frac{i}{ka}\sum_{j=1}^N e^{2ika j}
    \left(\frac{a}{a_j}\right)+ \sum_{j=2}^N \left(\frac{i}{ka}\right)^j
    \left\{ \sum_\sigma \left[ e^{2i ka \sigma_1} \left(\frac{a}{a_{\sigma_1}}\right) 
    \ldots \left(\frac{a}{a_{\sigma_j}}\right) \prod_{r=1}^{j-1}
    \left(1-e^{2ika(\sigma_{r+1}-\sigma_r)}\right)\right] \right\}
\end{align}
\end{subequations}
\end{widetext}
where for each $j$ the $\sum_\sigma$ means we are summing over the
$\binom{N}{j}$ combinations of size $j$ from the set $\{1,2,\ldots,N\}$
$\sigma=\{\sigma_1,\sigma_2,\ldots,\sigma_j\}$ with
$\sigma_1<\sigma_2<\ldots<\sigma_j$. The $r^R$ amplitude up to
a phase is obtained from
$r^L$ for the reverse chain. These latter formulae resemble the
equations for the band structure and eigenenergies of the closed system \cite{aro}.
In spite of their formidable aspect, Eqs.\eqref{eq:fool} are easy to program for
sequential calculations, providing the transmitivity and reflectivity of
the system with exact analytical expressions.

\begin{figure*}
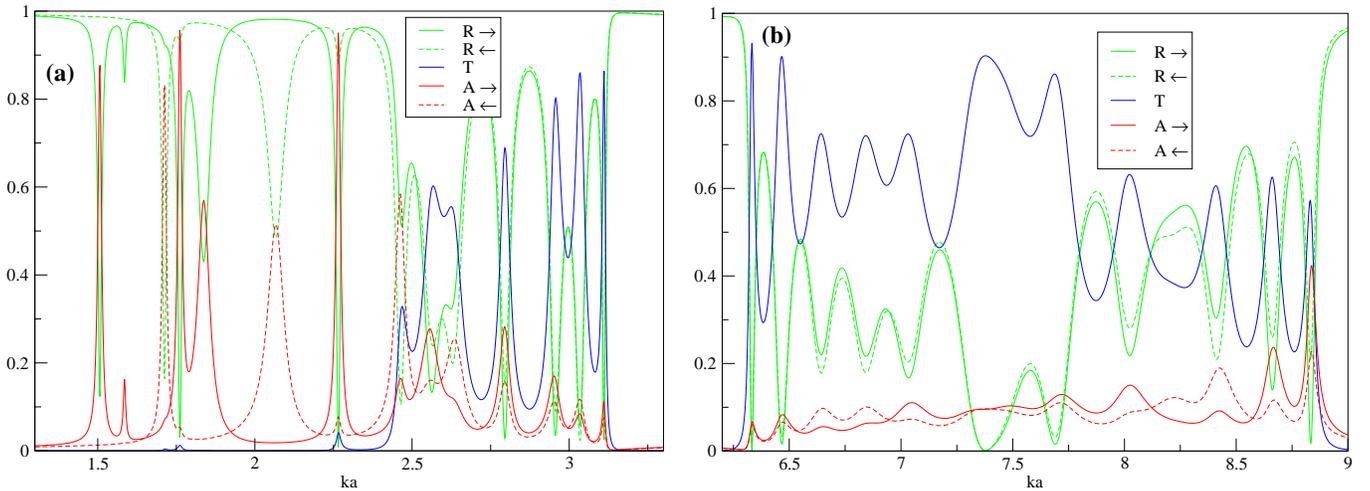

    \epsfig{file=fig1a.eps,width=.49\textwidth}\hfill
    \epsfig{file=fig1b.eps,width=.49\textwidth}
    \caption{(Color online) Scattering process for disordered arrays of $15$ deltas with
             complex couplings. The sequences of the real parts of the
       characteristic parameters are: (a) $\textrm{Re}(a/a_j)$: $3, 1, 2, 0.5,
        3, 2, 1, 3, 0.5, 4, 5,
        1, 2, 2, 3$ (b) $\textrm{Re}(a/a_j)$:$-1, -4, -3, -1, -2, -3,-4,$ $-1,
                -2,-3,-1,-4,-4,-2,-3$  . The imaginary part of each coupling has been
            chosen as $\textrm{Im}(a/a_j)=-0.01|\textrm{Re}(a/a_j)|$. The
    arrows in the legends mark the direction of incidence.}
    \label{fig:delran}
\end{figure*}
\begin{figure}
    \epsfig{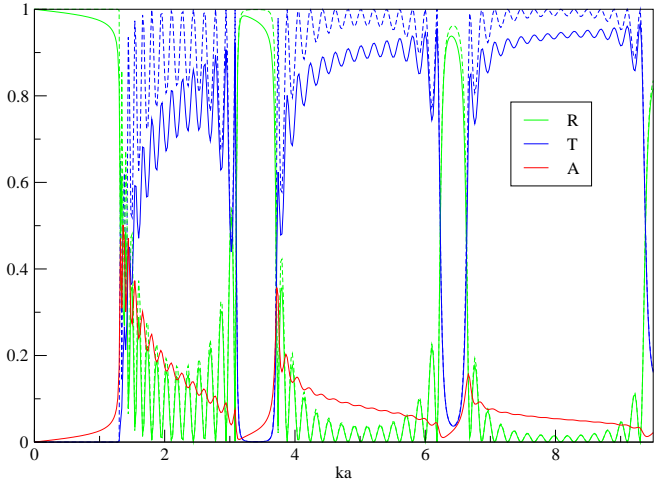}
    \caption{(Color online) Scattering and absorption probabilities for one-species delta
    chains with length $N=15$ and parameters $(a/a_1)=1.0$ (dashed lines)
    and $(a/a_1)=1.0-0.015i$ (solid lines).}
    \label{fig:delper}
\end{figure}
Let us now incorporate the dissipative processes that are always
present in real wires, causing energy
losses. We have modelled that effect by
including an imaginary part in the potential. In this
case the natural complex extension of our system consists in promoting the
delta couplings from real to complex, thus writing $(a/a_j)=r_j-i s_j$.
We also take $s_j>0$ for all $j$ in order to avoid anomalous
scattering.
The effect of including  complex couplings on the spectrum of an infinite
periodic delta 
chain has recently been studied in detail \cite{ptcer}. Let us see 
what happens in a chain with open boundaries. In Fig.\ref{fig:delper} the
usual scattering diagram is shown for a short periodic chain with real potentials. Including a
small imaginary part in the couplings we see how the transmission pattern
is altered with a non-negligible absorption that peaks at the incoming band
edges while the reflectivity is not noticeably 
changed. This tendency of the absorption term also 
appears when several species are included in the periodic array, and its
pattern  does not change much if
only some of the couplings are complexified.

When the array presents no ordering at all, the graph is quite
unpredictable and different configurations can be obtained. In
Fig.\ref{fig:delran}(a) a peaky spectrum with very sharp absorption
resonances is shown. The scattering process in this case is strongly dependent on the
direction incidence, as can be seen. On the other hand, smoother diagrams are
also possible in which the effect of the complex potential manifests
through an almost constant absorption background and a small change
depending upon 
the colliding side, like the one in Fig.\ref{fig:delran}(b).

This naive potential, apart from being exactly solvable, is powerful enough to
account for very different physical schemes, which makes it a very useful
bench-proof structure. 

%%%%%%%%%%%%%%%%%%%%%%%%%%%%%%%%%%%%%%%%%%%%%%%%%%%%%%%%%%%%%%%%%%%%%%%%%%%%%
\section{Atomic quantum wells}
    \label{sec:posch}

\begin{figure}
    \epsfig{file=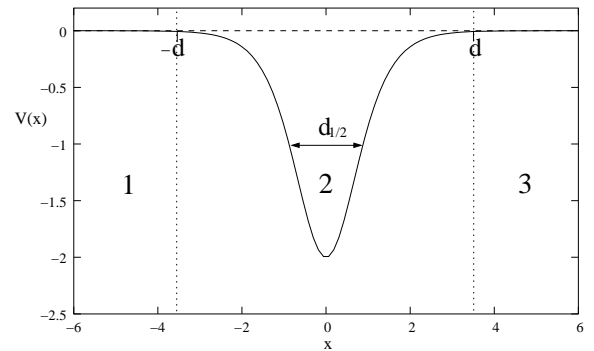,width=.9\columnwidth}
    \caption{P\"oschl-Teller potential (arbitrary units).}
    \label{fig:poteler}
\end{figure}

Let us go one step further and consider
a potential that resembles the profile of an atomic quantum well with
analytical solutions, the well-known P\"oschl-Teller potential hole.
It reads,
\begin{equation}
    V(x)=-\frac{\hbar^2}{2m} \alpha^2
    \frac{\lambda(\lambda-1)}{\cosh^2(\alpha x)} \quad \lambda >1 ,
    \label{eq:potencialpt}
\end{equation}
and it is shown in Fig.\ref{fig:poteler}.

The probability of asymptotic transmission 
is well known\cite{flugge},
\begin{equation}
    T_{PT}(k)=\frac{1}{1+p^2}\quad ;\quad p=\frac{\sin(\pi\lambda)}{\sinh(k\pi/\alpha)}.
    \label{eq:tpoteler}
\end{equation}
One characteristic feature of the P\"oschl-Teller hole is that it behaves as an
absolute transparent potential  for integer values of $\lambda$, as
can be seen from Eq.\eqref{eq:tpoteler}.
From the wave functions one can also obtain the asymptotic  
transmission matrix, 
\begin{equation}
    \mathbf{M}_{PT}(k)=\begin{pmatrix} 
        i e^{i\gamma} \sqrt{1+p^2} & -i p \\
        ip & -i e^{-i\gamma} \sqrt{1+p^2}
    \end{pmatrix}
    \label{eq:mpt}
\end{equation}
where $\gamma=2\arg\Gamma(ik/\alpha)-\arg\{\Gamma(\lambda+ik/\alpha)\Gamma(1-\lambda+ik/\alpha)\}$.
\begin{figure}
    \epsfig{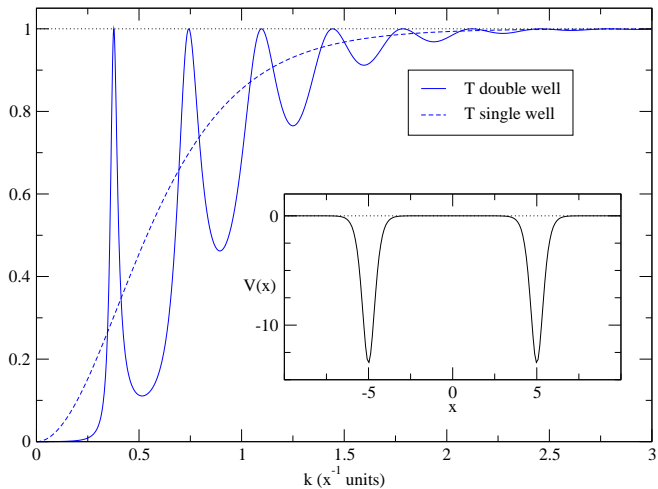}
    \caption{(Color online) Transmission through a double P\"oschl-Teller hole with
    parameters $\alpha_1=\alpha_2=2$ ($x^{-1}$ units), $\lambda_1=\lambda_2=2.4$,
    $d_1=d_2=5$ ($x$ units). The dashed line corresponds to a single potential hole. The
    inset shows the composite potential profile (arbitrary units).}
    \label{fig:postel1}
\end{figure}
\begin{figure}
    \epsfig{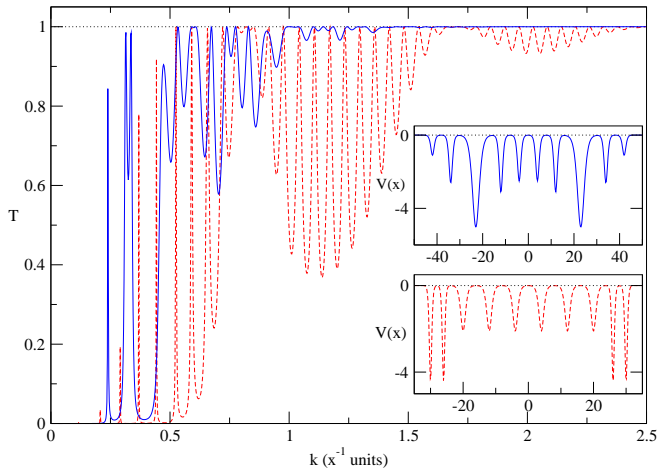}
    \caption{(Color online) Transmission patterns for two symmetric composite potentials of 
    ten units each. Their profiles are
    shown in the insets (arbitrary units). Parameters for the first five potentials of the sequences:  
    (solid line) $\alpha= 1,1,0.5,1,1$($x^{-1}$ units), $\lambda= 1.66,2.19,5.01,2.33,2.16$, 
    $d= 4,4,7,4,4$($x$ units) and (dashed line) $\alpha= 2,2,1,1,1$($x^{-1}$ units),
    $\lambda= 1.66,1.66,2.03,2.03,2.03$, $d = 2,2,4,4,4$($x$ units).}%    
    \label{fig:molecules}
\end{figure} 
Let us try to build a chain with these atomic units.
In order to do so, one has to include a sensible cut-off in
the potentials to ensure first that the wave function takes a proper form at the
junction regions and second that the resulting
 potential hole can still be described by a handy transmission
matrix, so that Eqs.\eqref{eq:comp} can be applied easily. The cut-off will be
placed at a distance $d$ from the center of the potential
(Fig.\ref{fig:poteler}). The wave function in the interval
$[-d,d]$ is $\Psi_2(x)=A_2 e(x)+B_2 o(x)$ where $e(x),o(x)$ are
the even and odd solutions respectively of the Schr\"odinger equation. Outside that interval the wave
function is assumed to be a superposition of the free particle solutions,
regions 1 and 3 in Fig.\ref{fig:poteler}.
The connection equations at the cut-off points lead to a relationship  between the
amplitudes of the wave function in sectors 1 and 3 in terms of the
values of  $e(x), o(x)$ and their spatial derivatives at $\pm
d$. Therefore, 
the distance $d$ must be such that the asymptotic form of the solutions of
the Schr\"odinger equation can be used at that point in order to ensure a
sensible transition to the free particle state and to obtain a
transmission matrix as simple as possible. The solutions
$e(x),o(x)$ as well as their asymptotic forms are found in Reference
\onlinecite{flugge}; nevertheless they are also reproduced in Appendix
\ref{ap:apenpoteler}.

After some algebra one finds the transmission matrix for the cut-off
version of the potential hole,
\begin{equation}
     \mathbf{M}(k)=\begin{pmatrix} 
        i e^{i(\gamma+2kd)} \sqrt{1+p^2} & -i p \\
        ip & -i e^{-i(\gamma+2kd)} \sqrt{1+p^2} 
    \end{pmatrix}.
    \label{eq:mspt}
\end{equation}
The matrix is the same as for the asymptotic case in Eq.\eqref{eq:mpt} 
plus an extra phase term in the diagonal elements that accounts for the
distance $2d$ during which the particle feels the effect of the potential. These
phases are the key quantities for the composition procedure since 
they will be responsible for the interference processes that produce the
transmission patterns. Due to the rapid decay of the P\"oschl-Teller
potential, the distance $d$ admits very reasonable values. In fact, we have
seen that for a sensibly wide range of the parameters $\alpha \in [0.1,3]$,
$\lambda\in[1,5]$ one can take as a minimum value for the cut-off distance $d_0\simeq
2 d_{1/2}=3.5/\alpha$, where $d_{1/2}$ is the half-width
(Fig.\ref{fig:poteler}). Taking $d\geq d_0$ the connection procedure works
really well, as we have checked in all cases by comparing the analytical composition
technique versus a high precision numerical integration of the
Schr\"odinger equation, obtaining an excellent degree of agreement. 

Composing two potential holes and applying Eqs.\eqref{eq:comp} together
with Eq.\eqref{eq:mspt} one finds for the transmission probability, using
the previously defined quantities $p$,$\gamma$
\begin{widetext}
\begin{equation}
    T_{2PT}(k)=\frac{1}{p_1^2 p_2^2+(1+p_1^2)(1+p_2^2)-2 p_1 p_2
    \sqrt{1+p_1^2}\sqrt{1+p_2^2}\cos(\gamma_1+ \gamma_2 +2k(d_1+d_2))},
    \label{eq:dpt}
\end{equation}
\end{widetext}
which is a handy expression that can hardly be obtained by trying to solve
the Schr\"odinger equation for the double potential hole. To our knowledge
this calculation has not been made before. 
Eq.\eqref{eq:dpt} clearly shows the interference effect depending on 
the distance $d_1+d_2$ between the centers of the holes. An example of
transmission is shown in Fig.\ref{fig:postel1}.

The composition procedure can be applied with a small number of atoms to
study the transmitivity of different potential profiles resembling 
molecular structures such as those in Fig.\ref{fig:molecules}. 
The transmission matrix \eqref{eq:mspt} can also be used to consider a
continuous disordered model in the form of a large chain of these potential holes
with random parameters. So far, in the literature only two kinds of
potentials have been used to build continuous disordered models, namely the
Dirac delta potential and the square well(barrier), due to their well known and easy
to manipulate transmission matrices. We recall the fact that handy
transmission matrices can be obtained for other potential profiles using 
reasonable approximations, such as the one described here.

The next step for our purpose is to consider dissipation in these
one-dimensional composite potentials.

%%%%%%%%%%%%%%%%%%%%%%%%%%%%%%%%%%%%%%%%%%%%%%%%%%%%%%%%%%%%%%%%%%%%%%%%%%%%%%%%%%%%%%%%
\section{Dissipative atomic quantum wells/barriers}
    \label{sec:Scarf}
\begin{figure}
    \epsfig{file=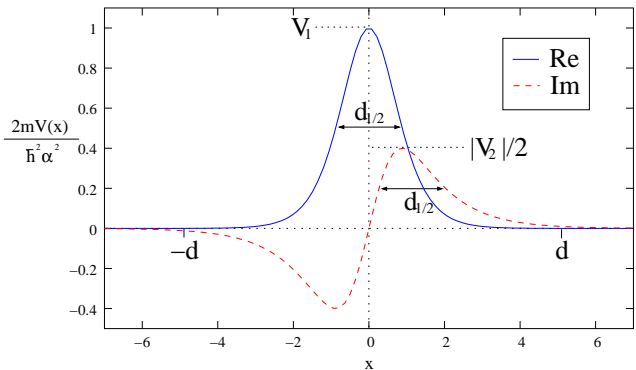,width=\columnwidth}
    \caption{Real and imaginary parts of the complex Scarf potential
    (arbitrary units).}
    \label{fig:potScarf}
\end{figure}

We shall consider the extension of the
P\"oschl-Teller potential  given by
the complexified Scarf potential,
\begin{equation}
    V(x)= \frac{\hbar^2 \alpha^2}{2m} \left(\frac{V_1}{\cosh^2(\alpha
    x)}+i\,V_2 \frac{\sinh(\alpha x)}{\cosh^2(\alpha x)}\right) 
    \label{eq:potScarf}
\end{equation}
with $V_1, V_2 \,\in\mathbb{R}$. It is a proper 
complex extension for two reasons: it admits
analytical solutions \cite{avinash} and its imaginary part is somehow proportional
to the derivative of the real potential. This latter criterion has been
considered in nuclear optical potentials to choose adequate complex
extensions. It seems reasonable to measure the strength of the 
dissipation processes
in terms of the ``density'' of the real interaction and therefore
writing an imaginary potential that is proportional to the spatial
derivative of the real one. The potential profile is shown in Fig.\ref{fig:potScarf}.

The Scarf potential has been extensively considered in the literature, 
mainly dealing with its discrete spectrum,
either in its real and complex forms, from the point of view of SUSY Quantum
 Mechanics \cite{susy} or focusing on its $\mathcal{PT}$-symmetric
form \cite{ahmed}. 

First, a detailed mathematical analysis of the
potential, regarding its scattering properties, 
must be made to discuss  some new features and some
assertions that have been made.

The left scattering amplitudes of the real Scarf potential have been
obtained in terms of complex Gamma functions \cite{avinash}. Recently, a considerable
simplification has been pointed out by Ahmed\cite{ahmed}. In fact, the
asymptotic transmitivity and reflectivity for the complex Scarf can be written as,
\begin{widetext}
\begin{align}
    T(k)&=
     \frac{\sinh^2(2\pi k/\alpha)}{\sinh^2(2\pi k/\alpha)+2\cosh(2\pi k/\alpha)\cosh(\pi g_+)
    \cosh(\pi g_-)+ \cosh^2(\pi g_+)+ \cosh^2(\pi g_-) } \label{eq:tScarf}\\
    R^L(k)&=\frac{\cosh^2(\pi g_+) e^{-2\pi k/\alpha}+\cosh^2(\pi
    g_-)e^{2\pi k/\alpha}  +2\cosh(\pi g_+)\cosh(\pi g_-)}{\sinh^2(2\pi k/\alpha)+2\cosh(2\pi k/\alpha)\cosh(\pi g_+)
    \cosh(\pi g_-)+ \cosh^2(\pi g_+)+ \cosh^2(\pi g_-)}
\end{align}
\end{widetext}
where $g_\pm=\sqrt{V_1\pm V_2-1/4}$ and $R^R(k)$ is recovered from $R^L(k)$
by interchanging $g_+$ and $g_-$ (which is equivalent to substituting
$V_2\rightarrow-V_2$ and therefore changing the direction of
incidence). These expressions derive from the asymptotic transmission
matrix, which is obtained here using the asymptotic form of the Schr\"odinger
equation solutions (Appendix \ref{ap:Scarf}), 
\begin{equation}
    M_{Scarf}(k)=\begin{pmatrix} i e^{i\varphi} \sqrt{1+s \overline{s}} & -i s \\
            i \,\overline{s} & -i e^{-i\varphi} \sqrt{1+s \overline{s}} \end{pmatrix}
    \label{eq:mScarf}
\end{equation}
where
\begin{align}
    s &=\frac{\cosh(\pi g_+)e^{\pi
        k/\alpha}+\cosh(\pi g_-)e^{-\pi k/\alpha}}{\sinh(2\pi k/\alpha)} \\
    \overline{s} &= s(g_+\rightleftarrows g_-)
\end{align}
and $\varphi=2\arg\{\Gamma(ik/\alpha)\Gamma(1/2+ik/\alpha)\}-\arg\{\Gamma(c +ik/\alpha)\Gamma(b 
 +ik/\alpha)\Gamma(1-c +ik/\alpha)\Gamma(1-b +ik/\alpha)\}$
with the definitions 
\begin{equation}
    c=\frac{1}{2}-\frac{i}{2}(g_+ - g_-)\quad;\quad
    b=\frac{1}{2}-\frac{i}{2}(g_+ + g_-).
    \label{eq:bcdef}
\end{equation}
It immediately follows from the transmission matrix that the absorption
probabilities read,
\begin{equation}
    A^L(k)= \frac{s\overline{s}-\overline{s}^2}{1+s\overline{s}}\quad;\quad
    A^R(k)= \frac{s\overline{s}-s^2}{1+s\overline{s}}.
    \label{eq:absor}
\end{equation}

\begin{figure}
    \epsfig{file=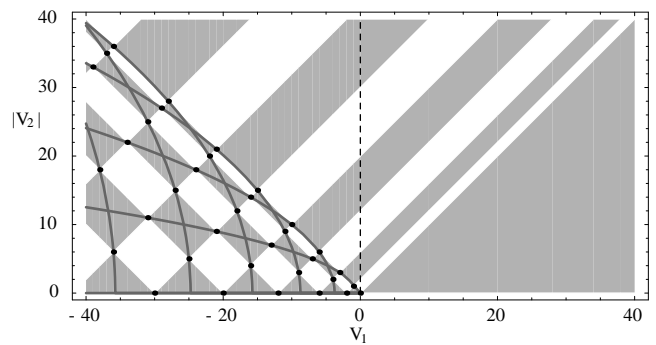,width=\columnwidth}
    \caption{Scattering diagram for the complex Scarf potential in terms of
the potential amplitudes. The physically acceptable ranges for $V_1,V_2$
correspond to the shaded zones. The curves are the inversion lines
given by Eq.\eqref{eq:invlines}. The black points mark the correlated
values of the amplitudes (Eqs.\eqref{eq:reson}) generating a fully
transparent behaviour.}
    \label{fig:chess}
\end{figure}

Unlike the complex delta potentials example this potential has some
drawbacks that must be carefully solved. Its imaginary part
is non-negative defined in its domain, which might cause anomalous
scattering. Only some values of $V_2$ will be physically acceptable. To
ensure that $T(k)\leq1 \quad\forall\, k$, it is clear from Eq.\eqref{eq:tScarf} that
the necessary and sufficient condition is $\cosh(\pi g_+)\cosh(\pi g_-)\geq
0$. The functions $g_+$, $g_-$ can be real or pure imaginary depending on
the values of $V_1$ and $V_2$. A detailed analysis of the conditions for
physical transmission is presented in Appendix \ref{ap:Scarf}. As a
summary, let us say that for $V_1>0$ (barriers) , the 
evaluation of the condition  translates into,
\begin{equation}
\begin{split}
     |V_2|\in& [0,V_1]\cup\\ &[2n(2n-1)+V_1,2n(2n+1)+V_1]
    \quad n\in\mathbb{Z}^+.
\end{split}
\end{equation}
For $V_1<0$ (wells) the situation becomes more complicated and the
result can only be expressed through several inequalities, each one
adding  a certain allowed range for $V_2$ (see Appendix \ref{ap:Scarf}). As an
example, in Table \ref{tab:v2} we show the compatible ranges of $V_2$ for a
few negative values
of $V_1$. One can trivially check the compatibility of the 
intervals presented for $V_2$ with the condition $T(k)\leq 1\quad \forall\,k$ by
plotting Eq.\eqref{eq:tScarf}.
In a two dimensional plot of $|V_2|$ vs. $V_1$, the
physical ranges for the transmission distribute as alternating 
fringes and a funny chessboard like pattern (Fig.\ref{fig:chess}).
\begin{table}
\begin{ruledtabular}
\begin{tabular}{ccc}
$V_1$ & $|V_2|$ & $|V_2|$ emissive \\\hline
$-0.5$ &  $[0,0.5]\cup[1.5,5.5]\cup[11.5,19.5]\ldots$ & $[0,0.5]$\\
$-1$   &  $[0,5]\cup[11,19]\cup[29,41]\ldots$ & $\emptyset$\\
$-2.4$ &  $[0,0.4]\cup[2.4,3.6]\cup[9.6,17.6]\ldots$ & $[0,0.4]\cup[2.4,2.569]$\\
$-3$   &  $[0,1]\cup\{3\}\cup[9,17]\ldots$ & $[0,1]$ \\
$-4$    & $[0,4]\cup[8,16]\cup[26,38]\ldots$ & $[3.606,4]$ \\
$-5$   &  $[0,1]\cup[3,5]\cup[7,15]\ldots$ & $[3,4.123]\cup[4.472,5]$
\end{tabular}
\end{ruledtabular}
\caption{Ranges of $V_2$ compatible with the condition $T(k)\leq 1
\quad \forall\,k$ for the complex Scarf potential for certain negative values of
$V_1$. The last column includes the intervals providing physical scattering
from the emissive side of the potential.}
\label{tab:v2}
\end{table}

One feature to emphasize according to the conditions given for acceptable
transmission is the fact that the number of permitted intervals for $V_2$
is infinite for any $V_1$, either positive(barrier) or negative(well), and
therefore there is no mathematical upper bound on
$|V_2|$($|V_2^{critical}|$) above which the transmission probability always
becomes unphysical, contrary to what has been reported recently \cite{ahmed}. From a
physical viewpoint of course, a sensible limitation must also be imposed on
$V_2$, usually $|V_2|\ll|V_1|$.

Let us see now what happens with the reflectivity. We assert that
for the values of $V_1$ and $V_2$ preserving a physical transmission, 
one of the reflectivities of the system remains physical (i.e. $R(k)\leq 1-T(k)\quad
\forall\,k$), left or right, 
depending on the particular values of $V_1,V_2$ (or equivalently, one of the
absorptions takes positive values for all $k$). The statement is easy to
prove from Eqs.\eqref{eq:absor} and more specifically reads: when $T(k)\leq
1\quad\forall\,k$ (i.e. $\cosh(\pi g_+)\cosh(\pi g_-)\geq0$), then 
\begin{equation}
    \begin{split}
     |\cosh(\pi g_-)|<|\cosh(\pi g_+)|&\Rightarrow A^L(k)\geq
    0\quad\forall\,k\\
    |\cosh(\pi g_-)|>|\cosh(\pi g_+)|&\Rightarrow A^R(k)\geq
    0\quad\forall\,k.
    \end{split}
    \label{eq:inversion}
\end{equation}
Considering $V_1>0$ and $V_2>0$ it is not hard to see that the first of
the above inequalities always holds. Therefore, in the case of a  potential barrier the
scattering is always physical from the absorptive side (trough of the
imaginary part), as has already been stressed \cite{ahmed}. More
interesting is the fact that this conclusion cannot be extended to the
case $V_1<0$ (well). In this case the physical scattering sometimes occurs
from the emissive side (peak of the imaginary part), producing smaller
absorption terms.
 In Table \ref{tab:v2} a few examples of
$V_2$ intervals providing physical scattering from the emissive side for
some potential wells are shown.

Another interesting feature that must be observed is that there exists a
set of correlated values of $V_1,V_2$ for which the complex Scarf potential
behaves as fully transparent. The condition for this to happen is from
Eq.\eqref{eq:tScarf} $\cosh(\pi g_+)=\cosh(\pi g_-)=0$. Thus, the main
requirement is that $g_+,g_-$ must be pure imaginary, yielding in this case
the transparency equations,
\begin{equation}
    \cos\left(\pi\sqrt{\frac{1}{4}-V_1\pm V_2}\right) = 0
\end{equation}
whose solution is,
\begin{subequations}
\label{eq:trans}
\begin{align}
    V_1 &= \frac{1}{4}-\frac{1}{8}\left[(2m+1)^2+(2n+1)^2\right]\\
    V_2 &= \frac{1}{8}\left[(2m+1)^2-(2n+1)^2\right]\quad m,n\in\mathbb{Z}.
\end{align}
\label{eq:reson}
\end{subequations}
It is worth noting that the transparencies only appear for potential wells
($V_1<0$). Considering the particular case $n=m$ one recovers the
P\"oschl-Teller resonances ($\lambda\in\mathbb{Z}$). In Table \ref{tab:trans} the first values of
Eqs.\eqref{eq:trans} are listed explicitly. 
\begin{table}
\begin{ruledtabular}
\begin{tabular}{ccccc}
     &  & $(-V_1$,$|V_2|)$ & & \\\hline
    $(1,1)$ & $(3,3)$ & $(6,6)$ & $(10,10)$ & $(15,15)$\\
    $[2,0]$ & $(4,2)$ & $(7,5)$ & $(11,9)$ & $(16,14)$\\
    $[6,0]$ & $(9,3)$ & $(13,7)$ & $(18,12)$ & $(24,18)$  \\
    $[12,0]$ & $(16,4)$ & $(21,9)$ & $(27,15)$ & $(34,22)$
\end{tabular}
\end{ruledtabular}
\caption{Some correlated values of $V_1,V_2$ producing the fully resonant
behaviour of the complex Scarf potential. The particular 
P\"oschl-Teller resonances are in square brackets.}
\label{tab:trans}
\end{table}
\begin{figure}
    \epsfig{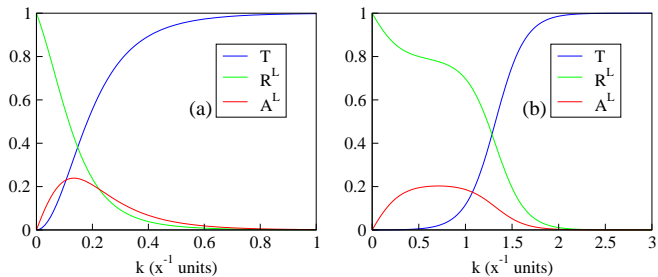}
    \caption{(Color online) Characteristic scattering patterns for (a) a complex Scarf well
$\alpha=1$($x^{-1}$ units),$V_1=-0.5,V_2=-0.4$ and (b) a complex Scarf
barrier $\alpha=1$($x^{-1}$ units),$V_1=2,V_2=0.1$.}
    \label{fig:plains}
\end{figure}

\begin{figure}
    \epsfig{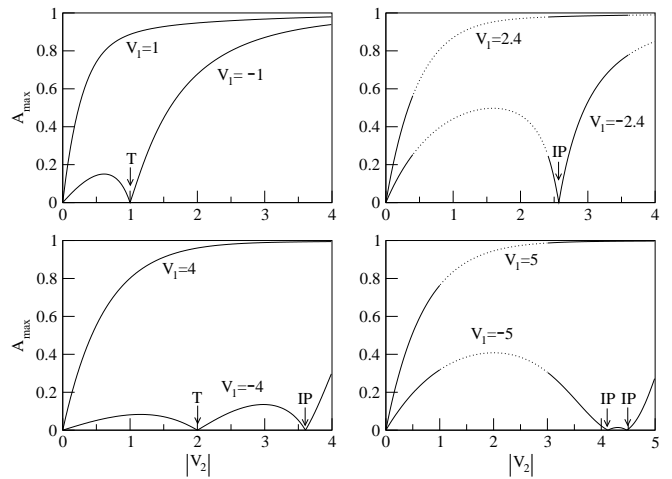}
    \caption{Maximum value of the absorption probability $A_{max}$ vs. $|V_2|$ for different values of
$V_1$. For each graph the upper(lower) curve corresponds to a barrier(well). A dotted
line is used in the forbidden values of $|V_2|$, nevertheless
the whole curve is shown for continuity. The allowed ranges of $|V_2|$ and
the position of the IP can be compared with the values in Table \ref{tab:v2}.}
    \label{fig:camel}
\end{figure}

The absorption obviously vanishes for all $k$ when considering these special resonant
values of the potential amplitudes. Surprisingly, there also exists another
set of non-trivial correlated values of $V_1,V_2$ for which the potential
is non-dissipative ($A(k)=0 \quad \forall\, k$) without being fully 
transparent. This set of values satisfies  $\cosh(\pi g_+)=\cosh(\pi
g_-)\neq 0$, as can be seen from Eqs.\eqref{eq:absor}. Non-trivial
solutions exist when $g_+,g_-\in\mathbb{C}$, yielding,
\begin{equation}
    |V_2|=n\sqrt{1-4V_1-4n^2}\quad n\in \mathbb{Z}^+.
    \label{eq:invlines}
\end{equation}
Let us also notice from Eqs.\eqref{eq:inversion} that these solutions 
are also the borders where the physical scattering changes 
from the absorptive side to the emissive side or vice versa. We shall refer
to these borders as inversion points (IP). Therefore, whenever we encounter an IP
we can say $A(k)=0 \quad \forall\, k$ without a fully transparent
behaviour, and hence a non-dissipative scattering process for all energies with a non-vanishing
imaginary part of the potential. Let us note that from
Eq.\eqref{eq:invlines} the IP only appears
in the case of Scarf potential wells and only for $|V_2|\leq 1/4-V_1$. In Fig.\ref{fig:plains} the
characteristic scattering probabilities are shown for a Scarf barrier and a
Scarf well, and in Fig.\ref{fig:camel} the maximum value of the physical
absorption is plotted versus $|V_2|$ for different values of $V_1$. When
$V_1$ is positive the absorption grows with the amplitude of the imaginary
part of the potential. On the other hand, for negative  $V_1$  a
strikingly different pattern arises, with transparencies (T) and inversion points
(IP) and the absorption does not increase monotonically with $|V_2|$. 
The whole behaviour of the scattering can be
clearly understood by building a two dimensional diagram $|V_2|$ vs. $V_1$ (Fig.\ref{fig:chess}),
including physically permitted ranges, inversion lines
and the points of transmission resonance.
 The complex Scarf potential shows two opposite faces to scattering, 
 namely barrier and well, and a much richer
structure in the latter case.
\begin{figure*}
    \centering
    \epsfig{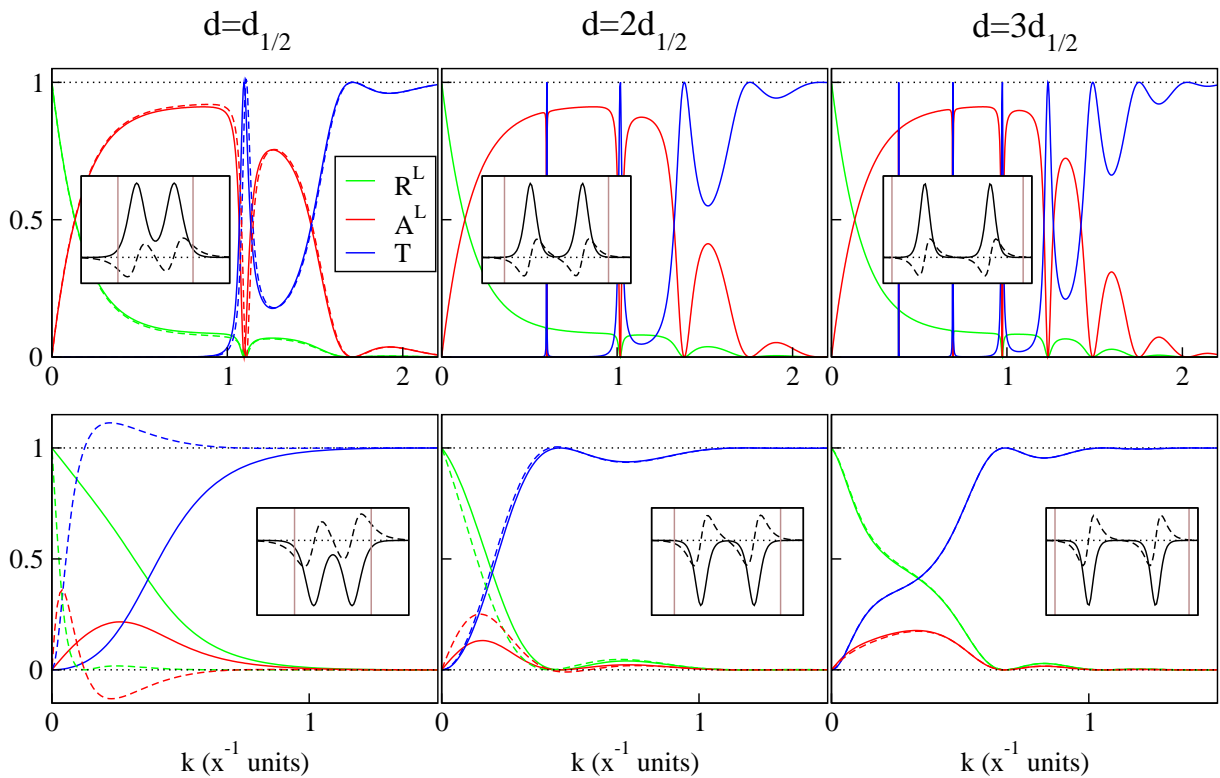}
    \caption{(Color online) Scattering probabilities for double Scarf  potentials with
parameters $\alpha=1$($x^{-1}$ units),$V_1=2,V_2=1$ for barriers and $\alpha=1$($x^{-1}$ units),$V_1=-4,V_2=3.1$ for wells. The
solid lines were obtained from the
analytical composition technique and the dashed lines correspond to
the exact numerical integration of the Schr\"odinger equation. The insets
show the potential profile (in arbitrary units) for each case: solid line for the real part and
dashed for the imaginary part. The vertical lines limit the portion of
the potential that the composition technique takes into account. For numerical
integration the whole potential profile was considered. Notice that
for low $d$ the exact integration may lead to unphysical scattering because
the conditions obtained for physical scattering are not valid for such $d$ values.
 As $d$ is increased an acceptable scattering is recovered and both
methods start converging. Convergence is reached faster in the case of
potential barriers where the analytical composition technique works
impressively well even for very low values of $d$.}
    \label{fig:transition}
\end{figure*}

After this detailed analysis of the peculiarities of the complex Scarf,
 that to our knowledge have not been reported before, let us
continue with our work on connecting several potentials to model
dissipative atomic chains. The procedure is the same as the one described
for the P\"oschl-Teller potential hole in Section \ref{sec:posch}. Only the
portion of the potential included in the interval $[-d,d]$ is taken
(Fig.\ref{fig:potScarf}). Making use of the asymptotic forms of the wave
functions, and after some algebra, one finds the transmission matrix for the
 complex Scarf potential with a symmetrical cut-off,
\begin{equation}
   M(k)=\begin{pmatrix} 
    i e^{i(\varphi+2kd)} \sqrt{1+s \overline{s}} & -i s \\
    i \,\overline{s} & -i e^{-i(\varphi+2kd)} \sqrt{1+s \overline{s}} 
    \end{pmatrix}.
\end{equation} 
The matrix proves to be the same as for the asymptotic case
(Eq.\eqref{eq:mScarf}), but with the extra phase $2kd$ in the diagonal
terms. It can be checked that the half-widths $d_{1/2}$ of both the real and imaginary
parts coincide and that the decay of the
imaginary part of the potential is always slower than that of the real part
(Fig.\ref{fig:potScarf}). This causes an increase in the minimum value of the cut-off distance $d_0$ 
with regard to the P\"oschl-Teller case. For sensible values of the
potential amplitudes we have found that considering $d_0\simeq 3d_{1/2}=5.3/\alpha$
is enough in most cases. In fact, this minimum value can be relaxed  in
the case of potential barriers ($V_1>0$), whereas for potential wells
taking $d$ below this value to apply the connection equations may sometimes
distort the results. The correct behaviour of the connection
procedure for $d\geq d_0$ can be observed in Fig.\ref{fig:transition} where
the scattering probabilities obtained upon integrating
the Schr\"odinger equation numerically are compared with those given by the analytical composition
technique, for two potential barriers and two
potential wells with different choices of the cut-off distance and with high
values for $|V_2|$. Notice how in
the barrier example convergence between the methods is completely
reached for  $d\simeq 2d_{1/2}$ whereas for potential wells a further step
 is needed because the convergence is much slower.

Once the connection of potentials has been successfully made, one should
ask which are the ranges of the potential amplitudes that provide an
acceptable physical scattering in this new framework. Analysis of this
issue is very non-trivial and quite complex analytically, but also very important because it determines
whether this model remains useful when considering atomic chains. First, 
in the case of two potentials we have observed that choosing 
each individual pair of amplitudes $(V_1,V_2)$ belonging to a physical 
range, and selecting the signs of the imaginary parts so that the physical
faces of both potentials point in the same direction, then an acceptable scattering for the composite potential can
always be recovered at least from one of the two possible orientations
(both physical faces to the right or to the left). In other words,
considering that the incident particle always collides  with the left side
(it comes from $-\infty$) and therefore orientating the individual physical
faces to the left, then at least one of the sequences $V_I(x)-V_{II}(x)$ or
$V_{II}(x)-V_I(x)$ gives an acceptable scattering for all energies. We have
checked this assertion for a broad variety of Scarf couples.
 For a higher number of potentials the situation becomes more
complex but a few pseudo-rules to obtain physical scattering can be deduced.
For an arbitrary chain we have found that in many cases the left
scattering remains physical as long as: the left scattering of each
individual potential is physical and the left scattering of each couple of
contiguous potentials is physical. This recipe seems completely true when
composing potential barriers only, whereas when wells are included it fails
in some situations, especially when several contiguous wells are surrounded
by barriers. Although at the beginning it may appear almost random to recover a
physical scattering from a large composition of Scarfs, following the given
advices it turns out to be more systematic. 

\begin{figure}
    \epsfig{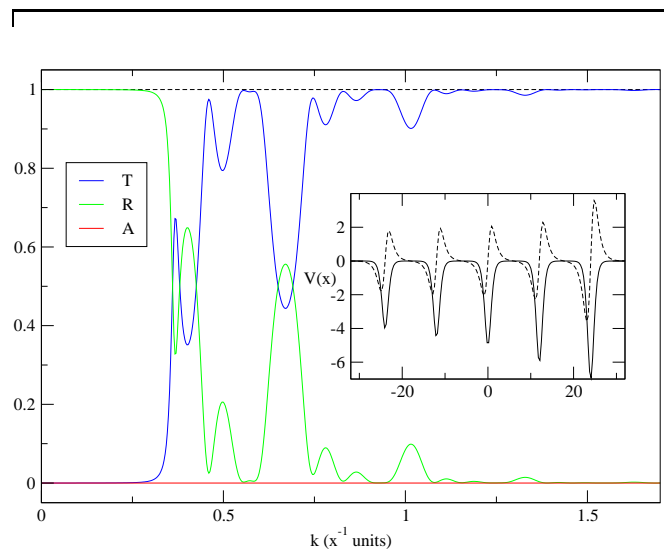}
    \caption{(Color online) Composition of Scarfs with different pair amplitudes
    describing IPs. Sequence with $\alpha=1$($x^{-1}$ units) and $(V_1,V_2)$: $(-4,\sqrt{13})$, $(-4.5,\sqrt{15})$,
    $(-5,\sqrt{17})$, $(-6,\sqrt{21})$, $(-7,2\sqrt{13})$. All the cut-off 
    distances have been taken equal to $d=6$($x$ units). The inset shows the
    potential profile with solid(dashed) line for the real(imaginary) part
    in arbitrary units. Notice that the
    composition remains non-dissipative.}
    \label{fig:IPs}
\end{figure}

Let us remember that the composition procedure, apart from being a powerful
tool for numerical calculations also provides analytical expressions for
the scattering probabilities, which of course adopt cumbersome forms for a large
number of potentials but are useful for obtaining simple expansions for
certain energetic regimes. Just as an example, the transmitivity and left
reflectivity for the
double Scarf read,
\begin{widetext}
\begin{align}
    T_{2Scarf}(k)&=\frac{1}{s_1^2 \overline{s}_2^2+(1+s_1\overline{s}_1)(1+s_2\overline{s}_2)-2 s_1 \overline{s}_2
    \sqrt{1+s_1\overline{s}_1}\sqrt{1+s_2\overline{s}_2}\cos(\varphi_1+ \varphi_2 +2k(d_1+d_2))}\\
    R^L_{2Scarf}(k)&=\frac{\overline{s}_1^2 (1+s_2\overline{s}_2)+\overline{s}_2^2 (1+s_1\overline{s}_1) 
    -2\overline{s}_1\overline{s}_2 \sqrt{1+s_1\overline{s}_1}\sqrt{1+s_2\overline{s}_2}\cos(\varphi_1+\varphi_2 +2k(d_1+d_2))}
    {s_1^2 \overline{s}_2^2+(1+s_1\overline{s}_1)(1+s_2\overline{s}_2)-2 s_1 \overline{s}_2
    \sqrt{1+s_1\overline{s}_1}\sqrt{1+s_2\overline{s}_2}\cos(\varphi_1+ \varphi_2 +2k(d_1+d_2))}
\end{align}
\end{widetext}    
making use of the previously defined terms $s,\overline{s}$ and
$\varphi$. 
One important feature of the formulae for the composite scattering probabilities
is the fact that they analytically account for the 
fully transparent behaviour of the whole structure as long as there is resonant
forward scattering of the individual potential units, as can be seen from
the latter equations and also from Eq.\eqref{eq:dpt} for the double
P\"oschl-Teller. Another curious feature arises when composing different
potentials whose amplitudes describe an IP. In this case the whole
structure remains non-dissipative (Fig.\ref{fig:IPs}). Moreover, the complex Scarfs at these
points behave completely as real potentials, providing an acceptable
scattering for the composition that is independent of the incidence
direction for any sequence of the individual Scarfs. 

\begin{figure}
    \epsfig{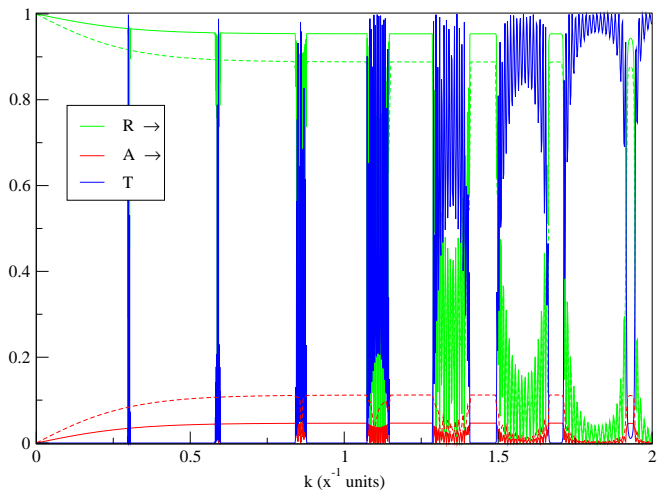}
    \caption{(Color online) Periodic chain of $20$ complex Scarfs with amplitudes
    $\alpha=1$($x^{-1}$ units), $V_1=2$, $V_2=0.02$ (solid lines) and $V_2=0.05$ (dashed lines) and
    cut-off distance $d=6$($x$ units). The
    transmitivity is not changed much by the small imaginary part of the potentials.}
    \label{fig:blochScarf}
\end{figure}

\begin{figure}
    \epsfig{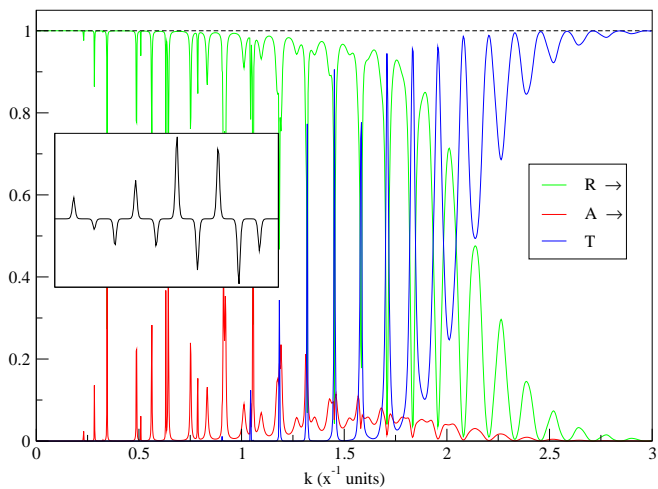}
    \caption{(Color online) Scattering probabilities for a $10$-Scarf structure with
    parameters $\alpha=1$($x^{-1}$ units) and $(V_1,V_2)$: $(1,0)$, $(-0.5,-0.005)$, $(-1.3,0.013)$,
    $(1.8,0.01)$, $(-1.3,0.013)$, $(4,0.04)$, $(-2.4,-0.024)$, $(3.5,0.04)$,
    $(-3.1,0.031)$, $(-1.5,0.04)$. The cut-off distance is $d=6$($x$ units) equal for
    all of them. The inset shows the real part of the potential
profile (arbitrary units).}
    \label{fig:scarmole}
\end{figure}

\begin{figure}
    \epsfig{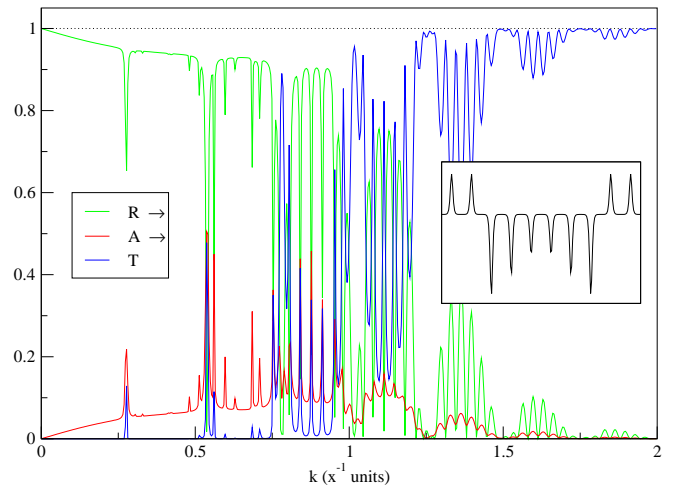}
    \caption{(Color online) Scattering probabilities for a $10$-Scarf structure with
    parameters $\alpha=1$($x^{-1}$ units) and $(V_1,V_2)$: $(1,0.04)$, $(1,0.05)$, $(-2,0)$,
    $(-1.5,0)$, $(-1,0)$, $(-1,0)$, $(-1.5,0)$, $(-2,0)$,
    $(1,0.1)$, $(1,0.1)$. The cut-off distance is $d=6$($x$ units) equal for
    all of them. The inset shows the real part of the potential profile
    (arbitrary units).}
    \label{fig:scaraggre}
\end{figure}

Considering larger Scarf chains with small imaginary parts of the
potentials, in the case of a periodic array we observe that the absorption
term remains flat over a wide range of  forbidden bands and oscillates
inside the permitted ones. The variations in the absorption are entirely balanced by
the reflectivity while the transmitivity is surprisingly not affected by
the presence of a small complex potential (Fig.\ref{fig:blochScarf}). This
behaviour contrasts strongly with  the complex delta potentials periodic
chain where the absorption was completely different and it was the
reflectivity that was little affected by the dissipation.
For an aperiodic sequence the situation is quite
different, as expected. In Fig.\ref{fig:scarmole} a type of molecular
aggregate is modelled with complex Scarfs. It exhibits a peaky absorption
spectrum and a strongly oscillating transmitivity with sharp resonances. In
Fig.\ref{fig:scaraggre} a symmetric atomic cluster has been considered in
which the dissipation only occurs at both ends.  
Different transmission and absorption configurations can be obtained by
building different structures. This shows the usefulness and versatility of
this model for being able to account for a variety of possible experimental
observations. 
\begin{figure*}
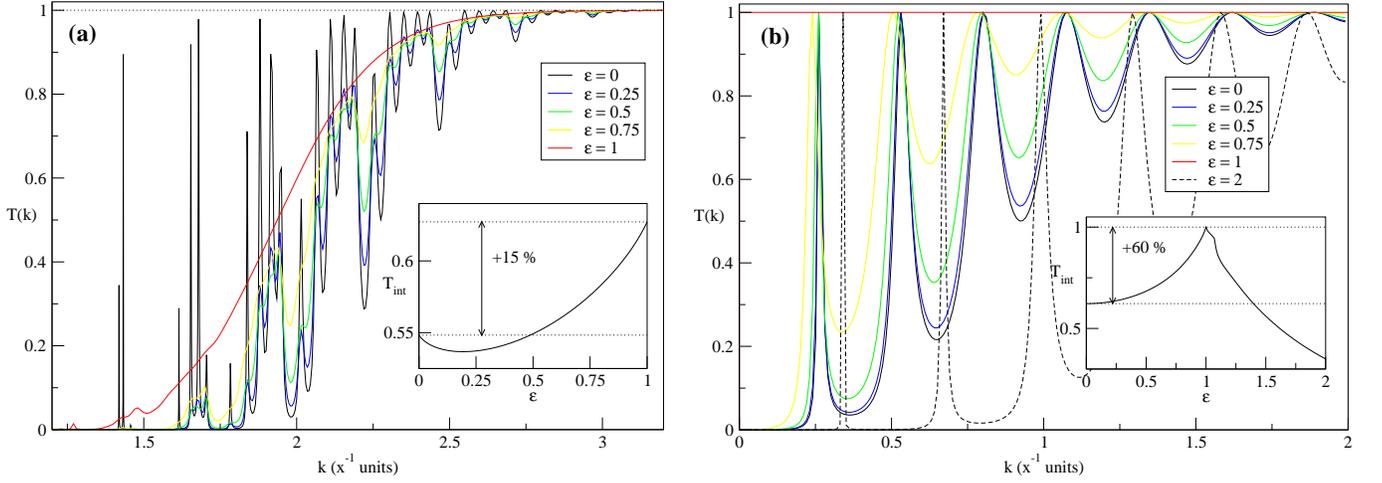

    \epsfig{file=fig15a.eps,width=.49\textwidth}\hfill
    \epsfig{file=fig15b.eps,width=0.49\textwidth}
    \caption{(Color online) Transmitivity for different strengths of the imaginary part of
    the potentials in (a) a chain of 6-Scarf barriers 
    $V_1=4,2,2.5,3,3.5,4$ and $\alpha=1$($x^{-1}$ units), 
    (b) a double Scarf well with $V_1=-1,-1$ and $\alpha=2$($x^{-1}$ units). The cut-off distances in all cases
    are equal to $d=6$($x$ units). Each imaginary amplitude reads $V_2=\varepsilon
    |V_1|$, except the first barrier in example (a) which is maintained
    real. The insets show the evolution of $T_{int}$ vs the strength of the
    imaginary amplitudes.}
    \label{fig:damping}
\end{figure*}

As a final exercise, two examples are included in Fig.\ref{fig:damping} showing the evolution of the
transmitivity of two different Scarf compositions as a function of the
imaginary part of the potential. The transmission patterns are plotted 
for different values of the parameter $\varepsilon=|V_2|/|V_1|$, which measures
the strength of the imaginary part. The
transmission efficiency is evaluated using an averaged transmitivity
$T_{int}=\int_{k_1}^{k_2} T(k) \,dk /(k_2-k_1)$ corresponding to the area enclosed by $T(k)$ per energy unit in a characteristic
energy range, namely the zone where $T$ evolves until it becomes saturated. The 
imaginary potential tends to smooth the transmission pattern in the first
example (corresponding to a sequence of barriers), causing a slight decrease in the averaged transmitivity for low
$\varepsilon$, although it is finally improved. For a double well a
different effect occurs. $T_{int}$ is always enhanced with increasing
$\varepsilon$ until it reaches a maximum, after which the transmitivity
falls with the imaginary potential. With these results in mind one could
speculate that these models might also be useful to describe the phase-breaking
inelastic scattering processes in atomic chains. Especially when one treats disordered
arrays where the break of the localization regime could arise as a result of
the loss of coherence due to inelastic collisions.

%%%%%%%%%%%%%%%%%%%%%%%%%%%%%%%%%%%%%%%%%%%%%%%%%%%%%%%%%%%%%%%%%%%%%%%%%%%%%%%%%%%%
\section{Concluding remarks}
  In this work we have used the transfer matrix formalism in the form of
the multiple collision technique to model dissipative scattering processes
by using complex potentials in various models of atomic chains. The
absorption probability has been rigorously included to recover unitarity for the
non-hermitian hamiltonians. 

New exact analytical expressions are given for the scattering amplitudes of an
arbitrary chain of delta potentials. The absorption effects arising by
promoting the delta couplings to the complex domain have been shown,
revealing the flexibility of this simple model to account for very
different physical schemes. 

Handy expressions for the transmission matrices of the P\"oschl-Teller and the complex Scarf
potential have been constructed in their asymptotic as well as their
cut-off versions for the first time. These latter matrices have made feasible, via the
composition technique, the assembly of an arbitrary number of potentials to
build atomic one-dimensional wires that can incorporate absorptive
processes. Different absorption configurations are presented in several 
examples that show the versatility of the model to account
for a variety of possible experimental observations.
The procedure developed is not only useful for numerical
calculations but also provides analytical formulae for the composite
scattering probabilities, which have not been obtained by other methods,
and whose significance has been checked by  numerical integrations of the
Schr\"odinger equation. 

From a more mathematical view, a complete and rigorous analysis of the scattering
properties of the complex Scarf potential has been carried out. The ranges
of physical transmission have been obtained and a group of novel features 
have arisen such as the presence of perfect transparencies and inversion points.

Apart from being 
able to include dissipation in the systems in a tractable way, the tools
and methods provided may have  direct
applicability for considering molecular aggregates \cite{molecule} and other
structures with explicit potential
profiles and also to build a new kind of continuous disordered models.

Future work seems promising since we are already in progress on the
assembly of the one-dimensional structures to quantum dots in order to 
analyze the effect of dissipation on the conductance. Our aim is also to
treat long disordered wires within this framework, and in particular to study
the applicability of these potentials to account for the loss of coherence due
to inelastic scattering process in the electronic localization regime.

%%%%%%%%%%%%%%%%%%%%%%%%%%%%%%%%%%%%%%%%%%%%%%%%%%%%%%%%%%%%%%%%%%%%%%%%
\begin{acknowledgments}
We thank E. Diez for several useful conversations. We acknowledge financial
support from DGICYT under contract BFM2002-02609.

\end{acknowledgments}

%%%%%%%%%%%%%%%%%%%%%%%%%%%%%%%%%%%%%%%%%%%%%%%%%%%%%%%%%%%%%%%%%%%%%%%%
\appendix
\section{Solutions of the P\"oschl-Teller potential}
    \label{ap:apenpoteler}
    The elementary positive energy solutions of the Schr\"odinger equation
with potential \eqref{eq:potencialpt} read,
\begin{subequations}
\begin{align}
    e(x) =&\cosh^\lambda(\alpha x)
    F\left(a,b,\frac{1}{2};-\sinh^2(\alpha x)\right)\\
    o(x) =&\sinh(\alpha x)\cosh^\lambda(\alpha x)\times \notag\\
    &F\left(a+\frac{1}{2},b+\frac{1}{2},\frac{3}{2};-\sinh^2(\alpha x)\right)
\end{align}
\end{subequations}
where $a=\frac{\lambda}{2}+i\frac{k}{2\alpha}$,
$b=\frac{\lambda}{2}-i\frac{k}{2\alpha}$ and $F(a,b,c;z)$ is the
Hypergeometric function. And their asymptotic forms can be written as,
\begin{subequations}
\begin{align}
    e(x) \rightarrow & \,m e^{-i\varphi} e^{-ik|x|} + m e^{i\varphi}
        e^{ik|x|}  \\
    o(x) \rightarrow & \,\textrm{sign}(x)\left( n e^{-i\theta} e^{-ik|x|} + n
    e^{i\theta} e^{ik|x|}\right) 
\end{align}
\end{subequations}
where 
\begin{subequations}
\begin{align}
   m e^{i\varphi} =&
    \frac{\sqrt{\pi}\,\Gamma\left(i\frac{k}{\alpha}\right)
    2^{-i\frac{k}{\alpha}}}
    {\Gamma\left(\frac{\lambda}{2}+i\frac{k}{2\alpha}\right) \Gamma\left(
    \frac{1-\lambda}{2}+i\frac{k}{2\alpha}\right)}\\ 
    n e^{i\theta} =&  \frac{\sqrt{\pi}\,\Gamma\left(i\frac{k}{\alpha}\right)
    2^{-i\frac{k}{\alpha}}}
    {2\,\Gamma\left(\frac{\lambda+1}{2}+i\frac{k}{2\alpha}\right) \Gamma\left(
    1-\frac{\lambda}{2}+i\frac{k}{2\alpha}\right)}.
\end{align}
\end{subequations}
 %%%%%%%%%%%%%%%%%%%%%%%%%%%%%%%%%%%%%%%%%%%%%%%%%%%%%%%%%%%%%%%%%%%%%%%
\section{Complex Scarf potential}
    \label{ap:Scarf}
\subsection{Scattering states}    
    The elementary positive energy solutions of the Schr\"odinger equation
with the complex Scarf potential \eqref{eq:potScarf} are,
\begin{subequations}
\begin{align}
    u_1(x) =& e^{-i\left(b-\frac{1}{2}\right)
    \arctan(\sinh(\alpha x))} \cosh^c(\alpha x) \times \notag \\
    & F\left(c+i\frac{k}{\alpha},c-i\frac{k}{\alpha},1-b+c;\frac{1}{2}+\frac{i}{2}\sinh(\alpha
    x)\right) \\
    u_2(x) =& e^{-i\left(b-\frac{1}{2}\right)
    \arctan(\sinh(\alpha x))} \cosh^c(\alpha x) \times \notag \\
    & \left(\frac{1}{2}+\frac{i}{2}\sinh(\alpha x)\right)^{b-c} \times \notag\\
    & F\left(b+i\frac{k}{\alpha},b-i\frac{k}{\alpha},1+b-c;\frac{1}{2}+\frac{i}{2}\sinh(\alpha
    x)\right)
\end{align}
\end{subequations}   
$F(a,b,c;z)$ being the hypergeometric function and using the $b,c$
definitions in \eqref{eq:bcdef}. The asymptotic limit
$x\rightarrow\pm\infty$ yields,
\begin{subequations}
\begin{align}
    u_1(x) & \rightarrow 2^c \left[\textrm{sign}(x)\cdot i\right] ^{\frac{1}{2}-b+c} \left(
    w_1 e^{-ik|x|} +w_2 e^{ik|x|}\right) \\
    u_2(x) & \rightarrow -\,2^c \left[-\textrm{sign}(x)\cdot i\right]^{\frac{3}{2}-b+c}\left( 
    q_1 e^{-ik|x|} +q_2 e^{ik|x|}\right)
\end{align}
\end{subequations}
where
\begin{subequations}
\begin{align}
        w_1 &=
    \frac{2^{2i\frac{k}{\alpha}}\Gamma\left(1-b+c\right)\Gamma\left(-2i\frac{k}{\alpha}\right)
    e^{-\textrm{sign}(x)\frac{\pi}{2}\frac{k}{\alpha}}
    }{\Gamma\left(c-i\frac{k}{\alpha}\right) \Gamma\left(1-b-i\frac{k}{\alpha}\right)}\\
    w_2 &=  \frac{2^{-2i\frac{k}{\alpha}}\Gamma\left(1-b+c\right)\Gamma\left(2i\frac{k}{\alpha}\right)
    e^{\textrm{sign}(x)\frac{\pi}{2}\frac{k}{\alpha}}
    }{\Gamma\left(c+i\frac{k}{\alpha}\right) \Gamma\left(1-b+i\frac{k}{\alpha}\right)}\\
        q_1 &= w_1\left(b\rightleftarrows c\right) \\
        q_2 &= w_2\left(b\rightleftarrows c\right).
\end{align}
\end{subequations}
\subsection{Physical transmission}%%%%%%%%%%%%%%%%%%%%
The condition for physical transmission is $\cosh(\pi g_+)\cosh(\pi g_-)\geq0$.
$V_2$ can be considered positive with no loss of generality since its change
in sign (which is equivalent to changing the side of incidence) does not affect the transmission.
With the definitions $X=|V_1|-V_2$ and $Y=|V_1|+V_2$, the
study can be easily carried out.
Considering $V_1>0$ the inequality translates into the permitted regions
\begin{equation}
 \forall \quad Y \begin{cases}
            X>0 \\
            -2n(2n+1)\leq X\leq -2n(2n-1)\quad n\in\mathbb{Z}^+
        \end{cases}
\end{equation}
which is clearly a sequence of allowed vertical fringes in the negative $X$
quadrant and the whole positive $X$ quadrant. This pattern will be same
but rotated $\pi/4$ clockwise when the change of variables is undone.
More specifically, in terms of the potential amplitudes the allowed intervals can be written
as
\begin{equation}
\begin{split}
     |V_2|\in& [0,V_1]\cup\\ &[2n(2n-1)+V_1,2n(2n+1)+V_1]
    \quad n\in\mathbb{Z}^+.
\end{split}
\end{equation}

In the case of $V_1<0$ a careful analysis leads to the following
cumbersome allowed sets
\begin{widetext}
\begin{align}
    X<0 & \Rightarrow 2n(2n-1)\leq Y \leq 2n(2n+1) \quad n=1,2,3,\ldots \\
    X>0 & \Rightarrow \begin{cases} 
            \big\{2n(2n-1)\leq X\leq 2n(2n+1) \big\}\bigcap
\big\{2m(2m-1)\leq Y\leq 2m(2m+1) \big\} \quad n,m=1,2,\ldots \\
             \big\{2j(2j+1)\leq X\leq 2j(2j+3)+2 \big\}\bigcap
\big\{2k(2k+1)\leq Y\leq 2k(2k+3)+2 \big\} \quad k,j=0,1,2,\ldots 
    \end{cases}
\end{align}
\end{widetext}
which are a set of allowed horizontal fringes in the negative $X$ quadrant and
a chessboard like structure for positive $X$. Undoing the change of
variables will mean a $\pi/4$ clockwise rotation followed by a reflection around
the vertical axis of this pattern to recover the negative
axis of $V_1$.
Solving these inequalities in terms of the potential amplitudes
gives rise to 
the following set of inequalities, each one assigning a certain
allowed interval for $|V_2|$ when fulfilled,
\begin{widetext}
for $n\in\mathbb{Z}^+$,
\begin{equation}
\begin{split}
        |V_1|\leq n(2n-1) &\Rightarrow
    \big[2n(2n-1)-|V_1|,2n(2n+1)-|V_1|\big] \\
        n(2n-1)\leq|V_1|\leq n(2n+1) &\Rightarrow \big[|V_1|,2n(2n+1)
    -|V_1|\big] 
\end{split}
\end{equation}
for $m,n\in\mathbb{Z}^+\quad m\geq n$,
\begin{equation}
\begin{split}
        m(2m-1)+n(2n-1)\leq |V_1| \leq m(2m-1)+n(2n+1) &\Rightarrow
    \big[2m(2m-1)-|V_1|, -2n(2n-1)+|V_1|\big]\\
    m(2m-1)+n(2n+1)\leq |V_1| \leq m(2m+1)+n(2n-1) &\Rightarrow
    \big[-2n(2n+1)+|V_1|, -2n(2n-1)+|V_1|\big]\\
    m(2m+1)+n(2n-1)\leq |V_1| \leq m(2m+1)+n(2n+1) &\Rightarrow
    \big[-2n(2n+1)+|V_1|, 2m(2m+1)-|V_1|\big]
\end{split}
\label{eq:ineq2}
\end{equation}
and for $j,k=0,1,2,\ldots \quad k\geq j$,
\begin{equation}
\begin{split}
     k(2k+1)+j(2j+1)\leq |V_1| \leq k(2k+1)+j(2j+3)+1 &\Rightarrow
    \big[2k(2k+1)-|V_1|, -2j(2j+1)+|V_1|\big]\\
    k(2k+1)+j(2j+3)+1\leq |V_1| \leq k(2k+3)+1+j(2j+1) &\Rightarrow
    \big[-2-2j(2j+3)+|V_1|, -2j(2j+1)+|V_1|\big]\\
    k(2k+3)+1+j(2j+1)\leq |V_1| \leq k(2k+3)+j(2j+3)+2 &\Rightarrow
    \big[-2-2j(2j+3)+|V_1|,2+ 2k(2k+3)-|V_1|\big].
\end{split}
\label{eq:ineq3}
\end{equation}
\end{widetext}
In the particular cases $m=n$ for Eqs.\eqref{eq:ineq2} and $k=j$ for
Eqs.\eqref{eq:ineq3} only the positive part of the allowed intervals must be
considered. The total physical range for $|V_2|$ comes from the union of the
different permitted intervals.

%%%%%%%%%%%%%%%%%%%%%%%%%%%%%%%%%%%%%%%%%%%%%%%%%%%%%%%%%%%%%%%%%%%%%%


\begin{thebibliography}{}
\bibitem{vieira} N. Agra\"{\i}t, C. Untiedt, G. Rubio-Bollinger and
    S. Vieira, Phys. Rev. Lett. \textbf{88}, 216803 (2002)
\bibitem{butti} M. B\"uttiker, Phys. Rev. B \textbf{32}, 1846 (1985);
    M. B\"uttiker, Phys. Rev. B \textbf{33}, 3020 (1986)
\bibitem{bundle} K. Maschke and M. Schreiber, Phys. Rev. B
    \textbf{44}, 3835 (1991); Phys. Rev. B \textbf{49}, 2295 (1994); R. Hey,
    K. Maschke and M. Schreiber, Phys. Rev. B \textbf{52}, 8184 (1995)
\bibitem{aro} J. M. Cerver\'o and A. Rodr\'{\i}guez, Eur. Phys. J. B
    \textbf{30}, 239 (2002); Eur. Phys. J. B
    \textbf{32}, 537 (2003)  
\bibitem{pataki} J. L. D'Amato and H. M. Pastawski, Phys. Rev. B
    \textbf{41}, 7411 (1990); 
\bibitem{kramer} G. Czycholl and B. Kramer, Solid State
    Commun. \textbf{32}, 945 (1979); D. J. Thouless and S. Kirkpatrick,
    J. Phys. C \textbf{14}, 235 (1981)
\bibitem{waves} V. Freilikher {\it et al}, Phys. Rev. Lett. \textbf{73},
    810 (1994); F. Delgado, J. G. Muga and A. Ruschhaupt, Phys. Rev. A
    \textbf{69}, 022106 (2004)
\bibitem{deych} L. I. Deych {\it et al}, Phys. Rev. B \textbf{64}, 024201
    (2001)
\bibitem{epscatt} C. Dembowski {\it et al}, Phys. Rev. Lett. \textbf{86},
    787 (2001); W. D. Heiss, Phys. Rev. E \textbf{61}, 929 (2000)
\bibitem{epcaos} W. D. Heiss, J. Phys. A \textbf{37}, 2455 (2004);
    Eur. Phys. J. D \textbf{7}, 1 (1999); W. D. Heiss and H. L. Harney,
    Eur. Phys. J. D \textbf{17}, 149 (2001); Eur. Phys. J. D \textbf{29},
    429 (2004)
\bibitem{nelson} N. Hatano and D. R. Nelson, Phys. Rev. Lett. \textbf{77},
    570 (1996); Phys. Rev. B \textbf{56}, 8651 (1997)
\bibitem{pts} C. M. Bender and S. Boettcher,
    Phys. Rev. Lett. \textbf{80}, 5243 (1998) ; C. M. Bender \textit{et al}, J. Phys. A \textbf{40},
    2201 (1999);R. M. Deb \textit{et al}, Phys. Lett. A \textbf{307}, 215
    (2003);
\bibitem{ptcer} J. M. Cerver\'o, Phys. Lett. A \textbf{317}, 26 (2003);
    J. M. Cerver\'o and A. Rodr\'{\i}guez, J. Phys. A. \textit{In Press}
    quant-ph/0312163 ; Z. Ahmed, Phys. Lett. A \textbf{286}, 231 (2001)
\bibitem{ptperio} C. Bender, G. V. Dunne and P. N. Meisinger, Phys. Lett. A
    \textbf{252}, 272 (1999); A. Khare and U. P. Sukhatme, Phys. Lett. A
    \textbf{324}, 406 (2004)
\bibitem{renorma} T. Khalil and J. Richert, J. Phys. A \textbf{37}, 4851
    (2004) 
\bibitem{report} J. G. Muga, J. P. Palao, B. Navarro and I. L. Egusquiza,
    Phys. Rep. \textbf{395}, 357 (2004)
\bibitem{harrison} P. Harrison, \textit{Quantum wells, wires and dots} John
Wiley and Sons Ltd, (2000)
\bibitem{zafarpra} Z. Ahmed, Phys. Rev. A \textbf{64}, 042716 (2001)
\bibitem{beam}  J. E. Beam, Am. J. Phys. \textbf{38}, 1395 (1970)
\bibitem{palao} J. P. Palao, J. G. Muga and R. Sala,
    Phys. Rev. Lett. \textbf{80} 5469 (1998)
\bibitem{adame} A. S\'anchez, E. Maci\'a and
    F. Dominguez-Adame, Phys. Rev. B \textbf{49}, 147 (1994) 
\bibitem{flugge} S. Fl\"ugge, \textit{Practical quantum mechanics},
    Springer-Verlag, Berlin (1970)
\bibitem{avinash} A. Khare and U. P. Sukhatme, J. Phys. A \textbf{21},
    L501 (1988);  J. W. Dabrowska, A. Khare and U. P. Sukhatme, J. Phys. A
    \textbf{21}, L195 (1988)
\bibitem{susy} J. M. Cerver\'o, Phys. Lett. A \textbf{153}, 1 (1991);
    G. L\'evai and M. Znojil, J. Phys. A \textbf{35}, 8793 (2002)
\bibitem{ahmed} Z. Ahmed, Phys. Lett. A \textbf{324}, 152 (2004)
\bibitem{molecule} R. Hey, F. Gagel, M. Schreiber and K. Maschke,
    Phys. Rev. B \textbf{55}, 4231 (1997)
\end{thebibliography}
\end{document}